\documentstyle[12pt]{article}\textwidth 16cm
\begin{document}
%
%
\begin{titlepage}
\begin{center}
{\Large\bf{ STRING TENSION AND THE GENERATION \\
\bigskip
OF THE CONFORMAL ANOMALY}}\\
\vspace{5em}
{\Large \bf  
A. Nicolaidis \footnote {nicolaid@ccf.auth.gr}},
{\Large \bf J. E. Paschalis \footnote {paschali@ccf.auth.gr}
 and 
P. I. Porfyriadis \footnote {porfyriadis@physics.auth.gr}}
\vspace{1em}\\
\vspace{2em}
Department of Theoretical Physics, University of
Thessaloniki, Thessaloniki 54006, Greece\\
\end{center}
\vspace{6em}
\begin{abstract}
The origin of the string conformal anomaly is studied in detail. 
We use a reformulated string Lagrangian which allows to consider 
the string tension $T_{0}$ as a small perturbation. The 
expansion parameter is the worldsheet speed of light c ,
which is proportional to $T_{0}$ . We examine carefully the interplay 
between a null (tensionless) string and a tensionful string which 
includes orders $ c^{2} $ and higher. The conformal algebra generated 
by the constraints is considered. At the quantum level the normal 
ordering provides a central charge proportional to $ c^{2} $. Thus 
it is clear that quantum null strings respect conformal invariance 
and it is the string tension which generates the conformal anomaly. 
\end{abstract}
\end{titlepage}
%
%

The most striking feature of quantum string models is their 
prediction of a critical spacetime dimension outside of which 
the quantum theory is problematical. Clearly it is quite important 
to understand how conformal invariance is broken by quantum effects. 
The whole issue becomes even more interesting by noticing the 
similarities between the canonical formulations of the string 
model and of general relativity. In both cases the Hamiltonian is 
weakly vanishing and the existing constraints reflect the underlying 
reparametrization invariance ( in two or four dimensions ) 
\cite{Hen}. It 
appears that we should address the problem within a framework which 
allows us to consider the string tension as a small perturbation. 
Such a formalism, the $c$ expansion, has been proposed by De Vega  
 and Nicolaidis in ref.[2]. The limit $T_{0}\rightarrow 0$ 
corresponds to the 
high energy limit of the string theory and it is also the appropriate 
limit for a string in strong gravitational field. The zeroth term in 
the $c$ expansion provides the  null string. The null string is a
tensionless string where each point of the string moves along a null 
geodesic \cite{Sch}. 
The quantization of a null string in de Sitter spacetime 
was studied in ref.[4]. No anomaly was found and it was conjectured 
that the dimensionful string tension lies at the origin of the 
conformal anomaly. We examine carefully here the quantum generation 
of the conformal anomaly and trace its explicit dependence upon the 
string tension. Aspects of this problem were addressed also in 
\cite{Lizzi}.

The starting point is the reformulated string Lagrangian \cite{VN}
\begin{equation}
L={1\over 4{\lambda}}[{\dot X^{\mu}}{\dot X^{\nu}}G_{\mu\nu}(X)
-c^2{X'^{\mu}}{X'^{\nu}}G_{\mu\nu}(X)] 
\end{equation}
where $c=2{\lambda}T_0$ is the world-sheet speed of light.
The string equations of motion read
\begin{equation}
{\ddot X^{\mu}}-c^2 X''^{\mu}
+{\Gamma^{\mu}_{\kappa\lambda}}({\dot X^{\kappa}}{\dot
X^{\lambda}}-c^2 X'^{\kappa} X'^{\lambda})=0
\end{equation}
supplemented by the constraints
\begin{eqnarray}
\Phi &\!\!\!\!=&\!\!\!\! \frac{1}{2}({\dot X^{\mu}}{\dot X^{\nu}}
G_{\mu\nu}+c^2{X'^{\mu}}{X'^{\nu}}G_{\mu\nu})=0 \\
\Psi &\!\!\!\!=&\!\!\!\! c {\dot X^{\mu}}X'^{\nu}G_{\mu\nu}=0 
\end{eqnarray}
Now we proceed to a series  expansion in powers of $c$ \cite{VN}. 
Writing
\begin{equation}
X^{\mu}({\sigma}, {\tau})=A^{\mu}({\sigma}, {\tau})
+c^2B^{\mu}({\sigma}, {\tau})+ \cdots 
\end{equation}
we obtain  the equations of 
motion for  the zeroth order term 
\begin{equation}
{\ddot A^{\mu}}+{\Gamma^{\mu}_{\kappa\nu}}
{\dot A^{\kappa}}{\dot A^{\nu}}=0 
\end{equation}
and the constraints
\begin{eqnarray}
\phi &\!\!\!\!=&\!\!\!\! \frac{1}{2}{\dot A^{\mu}}{\dot A^{\nu}}
G_{\mu\nu}=0 \\
\psi &\!\!\!\!=&\!\!\!\! 
c{\dot A^{\mu}}A'^{\nu}G_{\mu\nu}=0 
\end{eqnarray}
$A^{\mu}({\sigma}, {\tau})$ represents a tensionless string,
a collection of
massless particles moving independently along null geodesics.
For a string in Minkowski spacetime the null string solution 
is \cite{Sch}
\begin{equation}
A^{\mu}({\sigma}, {\tau}) = q^{\mu}(\sigma) + p^{\mu}(\sigma)\tau
\end{equation}
and it is obtained from the full string solution by keeping terms
constant  and linear in $c\tau$. String tension corrections appear 
when terms proportional to $ (c\tau)^{2} , (c\tau)^{3} 
\;.... $ are 
included.Thus $c$ expansion amounts to a short time expansion for the 
internal motion, while the overall string motion in the external 
geometry is treated exactly. In this sense $c$ expansion is an 
improvement over short time expansions proposed earlier \cite{SV,GSV}.       

Under the Poisson bracket
\begin{equation}
{\lbrace P^{\mu}(\sigma),X^{\nu}(\sigma') \rbrace}=\delta(\sigma -\sigma')
  G^{\mu\nu} 
\end{equation}
the constraints generate the conformal algebra ( we set 
$2 \lambda=1$)
\begin{eqnarray}
{\lbrace \Phi(\sigma),\Phi(\sigma') \rbrace} 
&\!\!\!\!=&\!\!\!\! c{\lbrack \Psi(\sigma)+\Psi(\sigma') \rbrack}
\delta'(\sigma -\sigma') \nonumber \\
{\lbrace \Phi(\sigma),\Psi(\sigma') \rbrace} 
&\!\!\!\!=&\!\!\!\! c{\lbrack \Phi(\sigma)+\Phi(\sigma') \rbrack}
\delta'(\sigma -\sigma') \\
{\lbrace \Psi(\sigma),\Psi(\sigma') \rbrace} 
&\!\!\!\!=&\!\!\!\! c{\lbrack \Psi(\sigma)+\Psi(\sigma') \rbrack}
\delta'(\sigma -\sigma') \nonumber 
\end{eqnarray}
Notice that the  algebra, eqs.(11) , is valid for any spacetime 
metric $G_{\mu \nu}$.
The constraints $\Psi$ and $\Phi$ generate two-dimensional 
reparametrizations of the string history $X^{\mu}(\sigma,\tau)$.
Indeed the generator
\begin{equation}
F=-\int ( \xi_{0}\Phi+\xi_{1}\Psi )d\sigma
\end{equation}
changes $X^{\mu}(\sigma,\tau)$ by
\begin{equation}
\delta X^{\mu}(\sigma,\tau)={\lbrace X^{\mu}(\sigma,\tau),F\rbrace}=
\xi_{0}\dot{X}^{\mu}(\sigma,\tau)+\xi_{1}X'^{\mu}(\sigma,\tau)
\end{equation}
The above transformation is identical to the one induced through 
infinitesimal changes of the world-sheet coordinates $\tau ,
\sigma$ by $\xi_0 , \xi_1$ respectively. 
Following the rule that null string results 
are obtained by dropping terms proportional to $ c^2 \; , \; c^3
\; ....$ , we obtain from eqs.(11), the algebra for a null string
\begin{eqnarray}
{\lbrace \phi(\sigma),\phi(\sigma') \rbrace} 
&\!\!\!\!=&\!\!\!\! 0 \nonumber \\
{\lbrace \phi(\sigma),\psi(\sigma') \rbrace} 
&\!\!\!\!=&\!\!\!\! c{\lbrack \phi(\sigma)+\phi(\sigma') \rbrack}
\delta'(\sigma -\sigma') \\
{\lbrace \psi(\sigma),\psi(\sigma') \rbrace} 
&\!\!\!\!=&\!\!\!\! c{\lbrack \psi(\sigma)+\psi(\sigma') \rbrack}
\delta'(\sigma -\sigma') \nonumber 
\end{eqnarray}
For open string, the Fourier components of the constraints 
\begin{eqnarray}
\Phi_n &\!\!\!\!=&\!\!\!\! \frac{2}{\pi}
\int_{0}^{\pi}\Phi(\sigma)\;cos\;n\sigma \;d\sigma \nonumber \\
&& \\
\Psi_n &\!\!\!\!=&\!\!\!\!  \frac{2}{\pi}
\int_{0}^{\pi}\Psi(\sigma)\;sin\; n\sigma\; d\sigma \nonumber  
\end{eqnarray} 
satisfy the algebra 
\begin{eqnarray}
{\lbrace \Phi_n,\Phi_m \rbrace} 
&\!\!\!\!=&\!\!\!\! - \frac{c}{\pi}(m-n)\Psi_{n+m} \nonumber\\
{\lbrace \Phi_n,\Psi_m \rbrace} 
&\!\!\!\!=&\!\!\!\! \frac{c}{\pi}(m-n)\Phi_{n+m} \\
{\lbrace \Psi_n,\Psi_m \rbrace} 
&\!\!\!\!=&\!\!\!\! \frac{c}{\pi}(m-n)\Psi_{n+m} \nonumber 
\end{eqnarray}
and similarly for the null string case, from eqs.(14), we obtain 
for the corresponding Fourier components
\begin{eqnarray}
{\lbrace \phi_n,\phi_m \rbrace} 
&\!\!\!\!=&\!\!\!\! 0 \nonumber\\
{\lbrace \phi_n,\psi_m \rbrace} 
&\!\!\!\!=&\!\!\!\! \frac{c}{\pi}(m-n)\phi_{n+m} \\
{\lbrace \psi_n,\psi_m \rbrace} 
&\!\!\!\!=&\!\!\!\! \frac{c}{\pi}(m-n)\psi_{n+m} \nonumber 
\end{eqnarray}

Moving from the classical to the quantum domain we replace the 
Poisson bracket by the commutator
\begin{equation}
{\lbrack P^{\mu}(\sigma),X^{\nu}(\sigma') \rbrack}=
        i\delta(\sigma -\sigma')  G^{\mu\nu} 
\end{equation}
The ordering of the operators offers the possibility for the 
emergence of central charges in the algebra. The appropriate 
ordering may be found by studying the quantum ground state. 
A string in Minkowski spacetime can be viewed as an ensemble of 
harmonic oscillators and normal ordering (annihilation operators 
to the right of creation operators) is the appropriate one. 
Writing for an open string in Minkowski spacetime 
\begin{equation}
X^{\mu}(\sigma,\tau)=q_{0}^{\mu}+\frac{1}{\pi}\Pi^{\mu}\tau+
\sum_{n=1}^{\infty}X_{n}^{\mu}(\tau)\;cos\;n\sigma 
\end{equation} 
we define the oscillator variables by
\begin{eqnarray}
\alpha_{n}^{\mu} 
&\!\!\!\!=&\!\!\!\! \frac{1}{2}(\dot{X}_{n}^{\mu}-
         incX_{n}^{\mu}) \nonumber\\
&& \\
 \alpha_{n}^{\mu*} 
&\!\!\!\!=&\!\!\!\! \frac{1}{2}(\dot{X}_{n}^{\mu}+
         incX_{n}^{\mu}) \nonumber
\end{eqnarray}
In terms of the new variables $ ( q_{0}^{\mu}\;,\;\Pi^{\mu}\;,\;
\alpha_{n}^{\mu}\;,\;\alpha_{n}^{\mu *} )$ the Fourier coefficients 
of the constraints take the form
\begin{eqnarray}
\Phi_{n}
&\!\!\!\!=&\!\!\!\! \frac{1}{\pi}\Pi( \alpha_{n}+
\alpha_{n}^{*})+\frac{1}{2}\sum_{m=1}^{n-1}(\alpha_{m}
\alpha_{n-m}+\alpha_{m}^{*}\alpha_{n-m}^{*})
+\sum_{m=1}^{\infty}(\alpha_{m}^{*}\alpha_{n+m}+
\alpha_{n+m}^{*}\alpha_{m}) \nonumber \\
&& \hspace{25em} ( n \neq 0 )\\
\Psi_{n}
&\!\!\!\!=&\!\!\!\! -\frac{i}{\pi}\Pi( \alpha_{n}-
\alpha_{n}^{*})-\frac{i}{2}\sum_{m=1}^{n-1}(\alpha_{m}
\alpha_{n-m}-\alpha_{m}^{*}\alpha_{n-m}^{*})
-i\sum_{m=1}^{\infty}(\alpha_{m}^{*}\alpha_{n+m}-
\alpha_{n+m}^{*}\alpha_{m}) \nonumber 
\end{eqnarray}
Eq.(18) provides the following commutators
\begin{eqnarray}
{\lbrack \Pi^{\mu},q_{0}^{\nu} \rbrack} 
                  &\!\!\!\!=&\!\!\!\! iG^{\mu\nu} \\
{\lbrack \alpha_{n}^{\mu},\alpha_{m}^{\nu *} \rbrack} 
      &\!\!\!\!=&\!\!\!\! -c\frac{n}{\pi}\delta_{nm}G^{\mu\nu} 
\end{eqnarray}
The quantum algebra generated by the constraints is evaluated using 
the normal ordering. After some lengthy calculations 
we obtain
\begin{eqnarray}
{\lbrack \Phi_n,\Phi_m \rbrack} 
&\!\!\!\!=&\!\!\!\! \frac{ic}{\pi}(n-m)\Psi_{n+m} \\
{\lbrack \Phi_n,\Psi_{m}^{*} \rbrack} 
&\!\!\!\!=&\!\!\!\! \frac{ic}{\pi}(n+m)\Phi_{n-m}+
\frac{id}{6\pi^2}c^2 (m^3-m)\delta_{nm} \\
{\lbrack \Psi_n,\Phi_{m}^{*} \rbrack} 
&\!\!\!\!=&\!\!\!\! -\frac{ic}{\pi}(n+m)\Phi_{n-m}-
\frac{id}{6\pi^2}c^2 (m^3-m)\delta_{nm} \\
{\lbrack \Psi_n,\Psi_m \rbrack} 
&\!\!\!\!=&\!\!\!\! - \frac{ic}{\pi}(n-m)\Psi_{n+m}  
\end{eqnarray}
The above quantum algebra is the main result of our work. 
The anomaly which appears in (25) and (26) , 
is of order $c^2$.  
 The anomaly originates from the reordering of the 
operators, each reordering providing a factor $c$ (eq.(23)).
The actual calculation indicates that the anomaly stems from the 
commutator $[\alpha_{k}\alpha_{n-k},\alpha_{l}^{*}
\alpha_{n-l}^{*}]$ (and its hermitean congugate), where 
eq.(23) is applied 
twice. Imagine a string with a vanishing small tension $T_0$. Our 
calculation shows that the anomaly vanishes also as $T_{0}^2$.
The fact that the string tension generates the conformal anomaly has been 
indicated also earlier \cite{BN}.

The transition from a quantum tensionful string to a quantum 
tensionless string is not a continuous process. In the case $T_{0}=0$, 
$\alpha^{\mu}_n$ and $\alpha^{\mu *}_n$ become identical and they 
commute. The appropriate ordering  then is to place the momentum 
operators $(P_n=\frac{1}{2\lambda}\dot{X_n})$ to the right of the 
position operators $(X_n)$. The following quantum algebra ( without 
any central charge) is obtained for the null string
\begin{eqnarray}
{\lbrack \phi_n,\phi_m \rbrack} 
&\!\!\!\!=&\!\!\!\! 0 \nonumber \\
{\lbrack \phi_n,\psi_{m}^{*} \rbrack} 
&\!\!\!\!=&\!\!\!\! \frac{ic}{\pi}(n+m)\phi_{n-m} \nonumber \\
{\lbrack \psi_n,\phi_{m}^{*} \rbrack} 
&\!\!\!\!=&\!\!\!\!- \frac{ic}{\pi}(n+m)\phi_{n-m} \\
{\lbrack \psi_n,\psi_m \rbrack} 
&\!\!\!\!=&\!\!\!\! -\frac{ic}{\pi}(n-m)\psi_{n+m}  \nonumber
\end{eqnarray}
It is important  to emphasize that the same algebra is obtained 
from the full quantum algebra, eqs. (24)-(27), following the rule 
we proposed, i.e. dropping terms $\; c^2 \;$ and higher. 
$\footnote{Regarding the quantization of a null string, an ordering 
prescription not implied by the physical problem itself, may lead 
to the emergence of an apparent quantum anomaly \cite{LG}}$

Summarizing, we studied the emergence of the string conformal anomaly.
We introduced a reformulated Lagrangian which allows to consider 
the limit $T_0 \rightarrow 0 $. We have shown that the conformal 
anomaly is proportional to $T^{2}_0$ and vanishes for a tensionless 
string. Throughout and at any stage, we are able to move from 
the full theory to the tensionless case (null string), 
via the correspondence rule we have established. Our explicit 
calculation demonstrates also the importance of the correct 
ordering of the quantum operators. 
We are familiar with the breaking of gauge 
invariance by a mass term and how this is cured through the Higgs 
mechanism. We may contemplate for a mechanism generating string 
tension and at the same time respecting the conformal symmetry.

\vspace{2em}
The authors wish to  acknowledge many useful discussions 
with Hector de Vega.
This work was supported in part by the General Secretariat of Research 
and Technology (Greece) and the EU program "Human Capital and 
Mobility".    
%
%

\end{document}